\documentclass[journal]{IEEEtranTCOM}

\usepackage{amsfonts}
\usepackage{amsmath}
\usepackage{amssymb}
\usepackage{graphicx}
\usepackage{dsfont}
\usepackage{algorithm}
\usepackage{algorithmic}

\normalsize

\begin{document}
%
\title{Improved Successive Cancellation Decoding of Polar Codes}
%
%
%

\author{Kai~Chen,
        Kai~Niu,
        and~Jia-Ru~Lin 
\thanks{K.~Chen, K.~Niu and J.~R.~Lin are with the Key Laboratory of Universal Wireless Communications, Ministry of Education, Beijing University of Posts and Telecommunications, Beijing 100876, China. E-mail: \{kaichen, niukai, jrlin\}@bupt.edu.cn}%
}


\maketitle

\begin{abstract}

As improved versions of successive cancellation (SC) decoding algorithm, successive cancellation list (SCL) decoding and successive cancellation stack (SCS) decoding are used to improve the finite-length performance of polar codes.
Unified descriptions of SC, SCL and SCS decoding algorithms are given as path searching procedures on the code tree of polar codes.
Combining the ideas of SCL and SCS, a new decoding algorithm named successive cancellation hybrid (SCH) is proposed, which can achieve a better trade-off between computational complexity and space complexity.
Further, to reduce the complexity, a pruning technique is proposed to avoid unnecessary path searching operations.
Performance and complexity analysis based on simulations show that, with proper configurations, all the three improved successive cancellation (ISC) decoding algorithms can have a performance very close to that of maximum-likelihood (ML) decoding with acceptable complexity.
Moreover, with the help of the proposed pruning technique, the complexities of ISC decoders can be very close to that of SC decoder in the moderate and high signal-to-noise ratio (SNR) regime.

\end{abstract}


%
\IEEEpeerreviewmaketitle

\section{Introduction}

\IEEEPARstart{P}{olar} codes, proposed by Ar{\i}kan \cite{Arikan}, are proved to achieve the symmetric capacities of the binary-input discrete memoryless channels (B-DMCs).
This capacity-achieving code family is based on a technique called channel polarization.
By performing the channel splitting and channel combining operations on independent copies of a given B-DMC, a set of synthesized binary-input channels can be obtained.
Let $I\left( W \right)$ denote the symmetric capacity of a B-DMC $W$.
It is proved in \cite{Arikan} that: with $N={{2}^{n}}$ uses of $W$, $n=1,2,\cdots $, when $N$ is large enough, it is possible to construct $N$ synthesized channels such that $N\left( 1-I\left( W \right) \right)$ of them are completely unreliable and $N I\left( W \right)$ of them are noiseless.
By transmitting free bits (called information bits) over the noiseless channels and transmitting a sequence of fixed bits (called frozen bits) over the others, polar codes can achieve the symmetric capacity under a successive cancellation (SC) decoder with both encoding and decoding complexity $O\left( N\log N \right)$.
In \cite{rate}, it is proved that the block error probability of polar code under SC decoding satisfies $P\left(N, R\right) \le {{2}^{-{{N}^{\beta }}}}$ for any $\beta <\frac{1}{2}$ when code length $N$ is large enough and code rate $R < I\left(W\right)$.
Furthermore, it was shown by Korada et al. \cite{characterization} that the error exponent $\beta $ can be arbitrarily close to 1 for large $N$ with a general construction using larger kernel matrices than the $2\times2$ matrix proposed by Ar{\i}kan.
To construct polar codes, the channel reliabilities can be calculated efficiently using Bhattacharyya parameters for binary-input erasure channels (BECs) \cite{Arikan}. But for channels other than BECs, density evolution is required \cite{DE}. More practical methods for calculating the channel reliabilities are discussed in \cite{construct} and \cite{construct_on}, and these techniques are extended to \mbox{$q$-ary} input channels \cite{constructq}.
The channel polarization phenomenon is believed to be universal in many other applications, such as parallel communications \cite{parallel} \cite{multilevel}, coded modulation systems \cite{pocm}, multiple access communications \cite{mac1} \cite{mac2}, source coding \cite{perform} \cite{source_coding1}, information secrecy \cite{security1} \cite{security2} and other settings.

Although polar codes have astonishing asymptotic performance, the finite-length performance of polar code under SC decoding is not satisfying. With the factor graph representation of polar codes, a belief propagation (BP) decoder is introduced by Ar{\i}an in \cite{Arikan}. And in \cite{perform}, Hussami et. al. show that BP decoder significantly can outperform SC decoder, and point out that, for channels other than BEC, the schedule of message passing in BP plays an important role. And they also show that the performance of BP decoder can be further improved by utilization of overcomplete factor graph representations over BEC. Unfortunately, due to the sensitivity of BP decoder to message-passing schedule, this is not realized on other channels.
In \cite{LP} a linear programming (LP) decoder is introduced without any schedule, and also, by using the overcomplete representations can improve the performance of LP decoder. But LP decoder cannot work on channels other than BEC.
Maximum likelihood (ML) decoders are implemented via Viterbi and BCJR algorithms on the codeword trellis  of polar codes \cite{ML}, but because of their high complexity, they can only work on very short code blocks.

Successive cancellation (SC) decoding of polar codes essentially shares the same idea with the recursive decoding of RM codes \cite{RMdec}.
Like the recursive decoders can be improved by using a list \cite{RMlist} or a stack \cite{RMstack}, SC can also be enhanced in the same way.

As an improved version of SC, successive cancellation list (SCL) decoding algorithm is introduced to approach the performance of maximum likelihood (ML) decoder with an acceptable complexity \cite{SCL:Tal}, \cite{SCL}.
And later, an other improved decoding algorithm based on SC named successive cancellation stack (SCS) decoding algorithm is proposed whose computational complexity will decrease with the increasing of signal-to-noise ratio (SNR) and can be very close to that of the SC decoding in the high SNR regime \cite{SCS}.
Compared with SCL, SCS will have a much lower computational complexity. But it comes at the price of much larger space complexity and it will fail to work when the stack is too small.
Combining the ideas of SCL and SCS, a new decoding algorithm named successive cancellation hybrid (SCH) is proposed in this paper, and it can achieve a better trade-off between computational complexity and space complexity.
In this paper, all the three improved SC decoding algorithms, SCL, SCS and SCH, are described under a unified manner of a path searching procedure on the code tree.
Further, to reduce the complexity, a pruning technique is proposed to avoid unnecessary path searching operations.

The remainder of the paper is organized as follows. Section \ref{section_preliminaries} reviews the basics of polar coding and describes the SC decoding algorithm as a path searching procedure on a code tree using \emph{a posteriori} probabilities (APPs) as metrics. Then the three improved successive cancellation (ISC) decoding algorithms and the pruning technique are introduced in section \ref{section_ISC}. Section \ref{section_simulations} provides the performance and complexity analysis based on the simulation results of polar codes under ISC decoders with different parameters. Finally, Section \ref{section_conclusion} concludes the paper.

\section{Preliminaries}
\label{section_preliminaries}

\subsection{Notation Convention}

In this paper, we use blackboard bold letters, such as $\mathbb{X}$ and $\mathbb{Y}$, to denote sets, and use $|\mathbb{X}|$ to denote the number of elements in $\mathbb{X}$.
We write the Cartesian product of $\mathbb{X}$ and $\mathbb{Y}$ as $\mathbb{X} \times \mathbb{Y}$, and write the $n$-th Cartesian power of $\mathbb{X}$ as ${\mathbb{X}}^n$.

We use calligraphic characters, such as $\mathcal{E}$ to denote a event. And let $\overline{\mathcal{E}}$ denote the event that $\mathcal{E}$ is not happened.

We use notation $v_1^N$ to denote a $N$-dimension vector $\left(v_1, v_2, \cdots, v_N \right)$ and $v_i^j$ to denote a subvector $\left(v_i, v_{i+1},\cdots, v_{j-1}, v_j\right)$ of $v_1^N$, $1\leq i,j \leq N$.
Particularly when $i>j$, $v_i^j$ is a vector with no elements in it and the empty vector is denoted by $\phi$.
We write $v_{1,o}^N$ to denote the subvector of $v_1^N$ with odd indices ($a_k: 1 \leq k \leq N$; $k$ is odd).
Similarly, we write $v_{1,e}^N$ to denote the subvector of $v_1^N$ with even indices ($a_k: 1 \leq k \leq N$; $k$ is even).
For example, for $v_1^4$, $v_{2}^3=(v_2,v_3)$, $v_{1,o}^4=(v_1,v_3)$ and $v_{1,e}^4=(v_2,v_4)$.
Further, given a index set $\mathbb{I}$, $v_{\mathbb{I}}$ denote the subvector of $v_1^N$ which consists of $v_i$s with $i \in \mathbb{I}$.

Only square matrices are involved in this paper, and they are denoted by bold letters.
The subscript of a matrix indicates its size, e.g. $\mathbf{F}_N$ represents a $N \times N$ matrix $\mathbf{F}$.
We write the Kronecker product of two matrices $\mathbf{F}$ and $\mathbf{G}$ as $\mathbf{F} \otimes \mathbf{G}$, and write the $n$-th Kronecker power of $\mathbf{F}$ as ${\mathbf{F}}^{\otimes n}$.

\subsection{Polar Codes}

Let $W: \mathbb{X} \to \mathbb{Y}$ denote a B-DMC with input alphabet $\mathbb{X}$ and output alphabet $\mathcal{Y}$. Since the input is binary, $\mathbb{X}=\left\{0,1\right\}$.
The channel transition probabilities are $W\left(y|x\right)$, $x\in \mathbb{X}$, $y\in \mathbb{Y}$.

For code length $N=2^n$, $n = 1,2,\cdots$, and information length $K$, i.e. code rate $R=K/N$, the polar coding over $W$ proposed by Ar{\i}kan can be described as follows:

After channel combining and splitting operations on $N$ independent uses of $W$, we get $N$ successive uses of synthesized binary input channels $W_N^{(i)}$, $i=1,2,\cdots,N$, with transition probabilities
\begin{equation}
\label{equ_polarized_channels}
     {W}_N^{(i)}(y_1^N, u_1^{i-1}|u_i)=\sum\limits_{u_{i+1}^{N} \in \mathbb{X}^{N-i}}{\frac{1}{2^{N-1}}{W}_N(y_1^N|x_1^N)}
\end{equation}
\noindent{where}
\begin{equation}
\label{equ_polarized_channels2}
    {W}_N(y_1^N|u_1^N)=\prod\limits_{{i}=1}^{N}{W(y_{i}|x_{i})}
\end{equation}
and the source block $u_1^N$ are supposed to be uniformly distributed in ${\left\{0,1\right\}}^{N}$.
Let ${P}_{e}\left(W_{N}^{\left(i\right)}\right)$ denote the probability of maximum-likelihood (ML) decision error of one transmission on $W_N^{\left(i\right)}$,
\begin{equation}
\label{equ_pe}
{{P}_{e}}\left( {W}_{N}^{\left( i \right)} \right)=
\sum\limits_{y_{1}^{N},u_{1}^{N}}{\frac{1}{2^N}{{W}_{N}}\left( y_{1}^{N}\left| u_{1}^{N} \right. \right)g_i\left( y_{1}^{N},u_{1}^{N} \right)}
\end{equation}
where $y_{1}^{N}\in{\mathbb{Y}^N}$, $u_{1}^{N} \in \{0, 1\}^N$ and the indicator function
\begin{equation}
g_i\left( y_{1}^{N},u_{1}^{N} \right)=
\begin{cases}
   1 & \text{if }\frac{{W}_{N}^{\left( i \right)}\left( y_{1}^{N},u_{1}^{i-1}\left| {{u}_{i}} \right. \right)}{{W}_{N}^{\left( i \right)}\left( y_{1}^{N},u_{1}^{i-1}\left| {{u}_{i}}\oplus 1 \right. \right)}\le 1  \\
   0 & \text{otherwise}  \\
\end{cases}
\end{equation}
and $\oplus$ is the module-$2$ addition.

The reliabilities of polarized channels $\left\{W_N^{(i)}\right\}$ are usually measured by (\ref{equ_pe}), and can be evaluated using Bhattacharyya parameters \cite{Arikan} for binary erasure channels (BECs) or density evolution \cite{DE} for other channels.

To transmit a binary message block of $K$ bits, the $K$ most reliable polarized channels $\left\{W_N^{(i)}\right\}$ with indices $i \in \mathbb{I}$ are picked out for carrying these information bits; and transmit a fixed bit sequence called frozen bits over the others.
The index set $\mathbb{I} \in \left\{1, 2, \cdots, N\right\}$ is called information set and $|\mathbb{I}|=K$. And the complement set of $\mathbb{I}$ is called frozen set and is denoted by $\mathbb{F}$.

Alternatively, the polar coding can be described as follow:
A binary source block $u_1^N$ which consists of $K$ information bits and $N-K$ frozen bits is mapped to a code block $x_1^N$ via $x_{1}^{N}=u_{1}^{N}{\mathbf{G}_{N}}$.
The matrix ${\mathbf{G}_{N}}={\mathbf{B}_{N}}\mathbf{F}_{2}^{\otimes n}$, where ${\mathbf{F}_{2}}=\left[ \begin{matrix}   1 & 0  \\   1 & 1  \\ \end{matrix} \right]$ and ${\mathbf{B}_{N}}$ is the bit-reversal permutation matrix.
The binary channel $x_{1}^{N}$ are then sent into channels which are obtained by $N$ independent uses of $W$.

\subsection{Successive Cancellation Decoding}

As mentioned in \cite{Arikan}, polar codes can be decoded by successive cancellation (SC) decoding algorithm.
Let $\hat{u}_1^N$ denote the estimate of the source block $u_1^N$.
After receiving $y_1^N$, the bits $\hat{u}_i$ are determined successively with index $i$ from $1$ to $N$ in the following way:
\begin{equation}
\label{equ_sc}
\hat{u}_i=
\begin{cases}
 {h_i}\left(y_1^N, \hat{u}_1^{i-1}\right) & i \in \mathbb{I}\\
 u_i & i \in \mathbb{F}\\
\end{cases}
\end{equation}
where
\begin{equation}
\label{equ_sc_h}
{h_i}\left(y_1^N, \hat{u}_1^{i-1}\right) =
\begin{cases}
 0 & \text{if }\tfrac{W_N^{(i)}\left(y_1^N, \hat{u}_1^{i-1} | 0\right)}{W_N^{(i)}\left(y_1^N, \hat{u}_1^{i-1} | 1\right)} \ge 1 \\
 1 & \text{otherwise}\\
\end{cases}
\end{equation}
The block error rate (BLER) of this SC decoding is upper bounded by
\begin{equation}
\label{equ_bler_sc}
P_{SC}\left(N,\mathbb{I}\right)\leq\sum\limits_{i\in\mathbb{I}}{P_e\left(\mathcal{W}_N^{(i)}\right)}
\end{equation}

\begin{figure}[t]
  \centering
  \includegraphics[width=0.8\columnwidth]{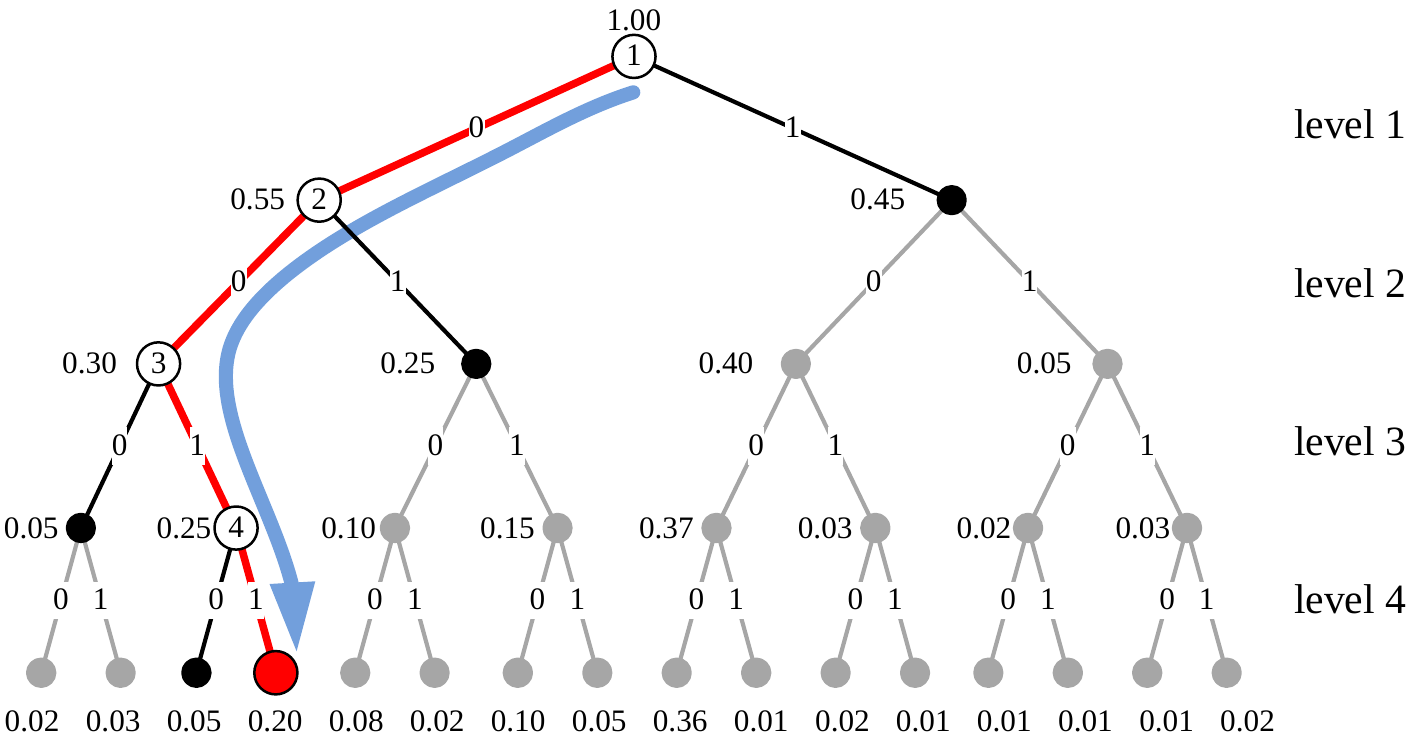}
  \caption{An example of code tree for code length $N=4$. The bold branches show a decoding path of SC with $\hat{u}_1^4=0011$.}
  \label{fig_tree}
\end{figure}

This successive decoding can be represented as a path searching process on a code tree.
For a polar code with code length $N$, the corresponding code tree $\mathbb{T}$ is a full binary tree.
More specifically, $\mathbb{T}$ can be represented as a 2-tuple $\left(\mathbb{V}, \mathbb{E}\right)$ where $\mathbb{V}$ and $\mathbb{E}$ denote the set of nodes and the set of edges respectively, $|\mathbb{V}|=2^{N+1}-1$, $|\mathbb{E}|=2^{N+1}-2$.
The depth of a node $v \in \mathbb{V}$ is the length of the path from the root to the node. The set of all nodes at a given depth $d$ is denoted by $\mathbb{V}_d$, $d=0, 1,\cdots, N$.
The root node has a depth of zero.
All the edges $e \in {\mathbb{E}}$ are partitioned into $N$ levels $\mathbb{E}_l$, $l=1,2,\cdots,N$, such that the edges in $\mathbb{E}_l$ incident with the nodes at depth $l-1$ and the nodes at depth $l$.
Except the nodes at the $N$-th depth $\mathbb{V}_N$, each $v \in \mathbb{V}_d$ has two descendants which belong to $\mathbb{V}_{d+1}$, and the two corresponding edges are label as $0$ and $1$ respectively.
The nodes $v \in \mathbb{V}_N$ are called leaf nodes.
Fig. \ref{fig_tree} gives a simple example of code tree with $N=4$.

A $i$-length decoding path $\left\{ {{e}_{1}},{{e}_{2}},...,{{e}_{i}} \right\}$ consists of $i$ edges, with $e_i \in \mathbb{E}_i$, $i \in \{1,2,\cdots, N\}$.
A vector $v_1^i$ is used to depict the above decoding path, where $v_i$ is corresponding to the binary label of edge $e_i$.
The reliability of a decoding path $v_1^i$ can be measured using \emph{a posteriori} probability
\begin{equation}
\label{equ_app}
P_N^{(i)}{\left(v_1^i|y_1^N\right)}=\frac{{W}_N^{(i)}(y_1^N, u_1^{i-1}|u_i)}{2P\left(y_1^N\right)}
\end{equation}

The APPs can be regarded as normalized versions of the channel transition probabilities defined in (\ref{equ_polarized_channels}).
The two kinds of probabilities are related by a multiplicative factor ${2P(y_1^N)}$.
By eliminating the factor, the APPs take values in a more stable range, and all the decoding paths with the same lengths have the sum probability equals to one, i.e.
\begin{equation}
\sum \limits_{v_1^i \in \{0,1\}^i}{P_N^{(i)}{\left(v_1^i|y_1^N\right)}}=1
\end{equation}
This property will help in understanding the path searching procedure in the code tree and is more suitable for hardware implementation.

Similar to the recursive expressions of (\ref{equ_polarized_channels}) given in \cite{Arikan}, the APPs can also be calculated recursively.
For any $n\ge 0$, $N={{2}^{n}}$, $1\le i\le N$,
    \begin{equation}
    \label{equ_app_recursive1}
        \begin{aligned}
        & P_{2N}^{\left( 2i-1 \right)}\left( v_{1}^{2i-1}\left| y_{1}^{2N} \right. \right) \qquad \qquad \qquad \qquad \qquad\\
        & =\sum\limits_{{{v}_{2i}\in\{0,1\}}}{P_{N}^{\left( i \right)}\left( v_{1,o}^{2i}\oplus v_{1,e}^{2i}\left| y_{1}^{N} \right. \right)
        \cdot P_{N}^{\left( i \right)}\left( v_{1,e}^{2i}\left| y_{N+1}^{2N} \right. \right)}
        \end{aligned}
    \end{equation}
    \begin{equation}
    \label{equ_app_recursive2}
    \begin{aligned}
    & P_{2N}^{\left( 2i \right)}\left( v_{1}^{2i}\left| y_{1}^{2N} \right. \right) \qquad \qquad \qquad \qquad \qquad \qquad\\
    & =P_{N}^{\left( i \right)}\left( v_{1,o}^{2i}\oplus v_{1,e}^{2i}\left| y_{1}^{N} \right. \right)
    \cdot P_{N}^{\left( i \right)}\left( v_{1,e}^{2i}\left| y_{N+1}^{2N} \right. \right)\qquad \quad
    \end{aligned}
    \end{equation}

SC decoding can be seen as a greedy search algorithm on the code tree.
In each level, only the one of two edges with larger probability is selected for further processing.

The red bold edges in Fig. \ref{fig_tree} shows the SC decoding path.
The number written next to each of the nodes provides the APP metric of the decoding path from the root to that node.
The nodes which are extended during the SC decoding procedure are represented by the numbered circles, and the corresponding numbers indicate the processing order.
The black circles represent the nodes which are visited (whose APP metric is calculated) but failed in competition for further exploring.
And the gray ones are those which are not visited during the searching process.
In the example, four times of calculations of equation (\ref{equ_app}) are required, one for each level.
However, the decoding path is not guaranteed to be the most probable one.
As shown in the example, the one labeled $1000$ has the largest probability of all the $N$-length paths, but it failed in the competition at the first level.

For further practical considerations, we use the logarithmic APPs as the path metrics:
\begin{equation}
\label{equ_metric_info_or_froz}
M_{N}^{\left( i \right)}\left( v_{1}^{i}\left| y_{1}^{N} \right. \right)=
\begin{cases}
\log P_{N}^{\left( i \right)}\left( v_{1}^{i}\left| y_{1}^{N} \right. \right)& i\in \mathbb{I} \\
M_{N}^{\left( i \right)}\left( v_{1}^{i-1}\left| y_{1}^{N} \right. \right) & i\in {{\mathbb{F}}} \\
\end{cases}
\end{equation}
For $i\in \mathbb{I}$, the path metric can be recursively calculated as
\begin{equation}
\label{equ_metric_odd}
\begin{aligned}
& M_{N}^{\left( 2i-1 \right)}\left( v_{1}^{2i-1}\left| y_{1}^{N} \right. \right)= \\
& {{\max }^{*}}\left\{ M_{N/2}^{\left( i \right)}\left( v_{1,o}^{2i}\oplus {{v}'}_{1,e}^{2i}\left| y_{1}^{N/2} \right. \right)+M_{N/2}^{\left( i \right)}\left( {{v}'}_{1,e}^{2i}\left| y_{N/2+1}^{N} \right. \right), \right. \\
& \quad \left. M_{N/2}^{\left( i \right)}\left( v_{1,o}^{2i}\oplus \bar{{v}'}_{1,e}^{2i}\left| y_{1}^{N/2} \right. \right)+M_{N/2}^{\left( i \right)}\left( \bar{{v}'}_{1,e}^{2i}\left| y_{N/2+1}^{N} \right. \right) \right\} \\
\end{aligned}
\end{equation}
\noindent and
\begin{equation}
\label{equ_metric_even}
\begin{aligned}
& M_{N}^{\left( 2i \right)}\left( v_{1}^{2i}\left| y_{1}^{N} \right. \right)= \\
& \quad M_{N/2}^{\left( i \right)}\left( v_{1,o}^{2i}\oplus v_{1,e}^{2i}\left| y_{1}^{N/2} \right. \right)+M_{N/2}^{\left( i \right)}\left( v_{1,e}^{2i}\left| y_{N/2+1}^{N} \right. \right) \\
\end{aligned}
\end{equation}

\noindent where function ${{\max }^{*}}\left( a,b \right)=\max \left( a,b \right)+\log \left( 1+e^{-\left| a-b \right|} \right)$ is the Jacobian logarithm and ${{v}'}_{1,e}^{2i}=\left\{ {{v}_{2}},{{v}_{4}},\cdots ,{{v}_{2i}}=0 \right\}$, $\bar{{v}'}_{1,e}^{2i}=\left\{ {{v}_{2}},{{v}_{4}},\cdots ,{{v}_{2i}}=1 \right\}$.

Then, the decision function of SC in (\ref{equ_sc_h}) is rewritten as
\begin{equation}
\label{equ_sc_h2}
{h_i}\left(y_1^N, \hat{u}_1^{i-1}\right) =
\begin{cases}
 0 & \text{if } \frac{M_{N}^{(i)}({{u}_{i}}=0,\hat{u}_{1}^{i-1}|y_{1}^{N})}{M_{N}^{(i)}({{u}_{i}}=1,\hat{u}_{1}^{i-1}|y_{1}^{N})}\ge 1 \\
 1 & \text{otherwise}\\
\end{cases}
\end{equation}


Using the space-efficient structure \cite{SCL:Tal} to implement a SC decoder, the time and space complexity are $O(N \log N)$ and $O(N)$ respectively.

\section{Improved Successive Cancellation Decoding}
\label{section_ISC}

The performance of SC is limited by the bit-by-bit decoding strategy. Since whenever a bit is wrongly determined, there is no chance to correct it in the future decoding procedure.

Theoretically, the performance of the maximum \emph{a posteriori} probability (MAP) decoding (or equivalently ML decoding, since the inputs are assumed to be uniformly distributed) can be achieved by traversing all the $N$-length decoding paths in the code tree $\mathbb{T}$.
But this brute-force traverse takes exponential complexity and is difficult to be implemented.

Two improved decoding algorithms called successive cancellation list (SCL) decoding and successive cancellation stack (SCS) are proposed in \cite{SCL:Tal} \cite{SCL} and \cite{SCS}.
Both of these two algorithms allow more than one edge to be explored in each level of the code tree.
During the SCL(SCS) decoding, a bunch of candidate paths will be obtained and stored in a list(stack).
Since for every single candidate path, the metric calculations and bit determinations are still performed bit-by-bit successively, SCL and SCS can be regarded as two improved versions of conventional SC decoding.

In this section, we will restate SCL and SCS under a unified framework with the help of APP metrics and the code tree representations.
Then to overcome the own shortages of SCL and SCS, a new hybrid decoding algorithm named successive cancellation hybrid (SCH) decoding is proposed.
Furthermore, to reducing the computational complexities, we propose a pruning technique to eliminate the unnecessary calculations during the path searching procedure on the code tree.

\subsection{Successive Cancellation List Decoding}

\begin{figure}[!t]
  \centering
  \includegraphics[width=0.8\columnwidth]{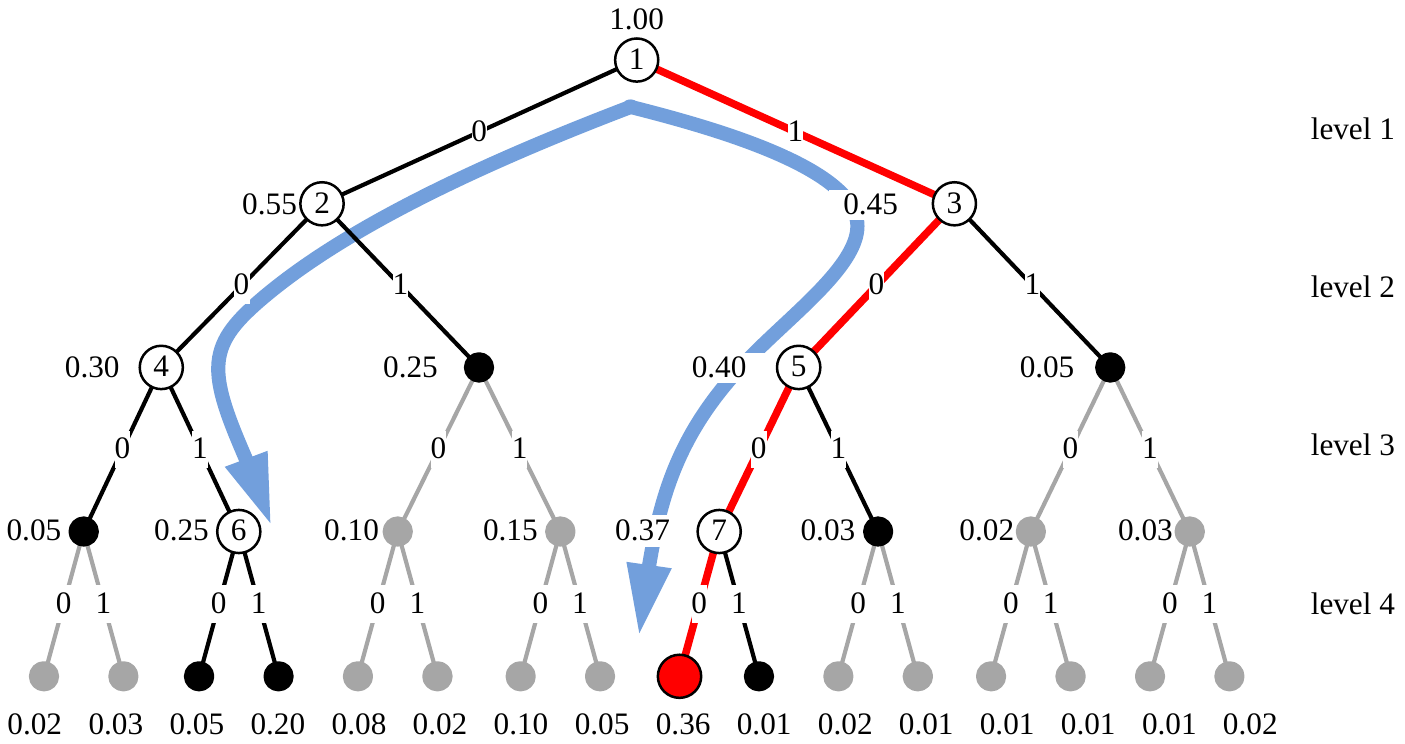}
  \caption{An example of SCL decoding with searching width $L=2$.}
  \label{fig_scl}
\end{figure}

As an enhanced version of SC, the successive cancellation list (SCL) decoder \cite{SCL:Tal} \cite{SCL} searches level-by-level on the code tree, which is just the same with SC.
However, unlike SC where only one path is reserved after processing at each level, SCL allows at most $L$ candidate paths to be further explored at the next level.

SCL can be regarded as a breadth-first searching on the code tree ${T}$ with a searching width $L$.
At each level, SCL doubles the number of candidates by appending a bit $0$ or a bit $1$ to each of the candidate paths, and then selects at most $L$ ones with largest metrics and stores them in a list for further processing at the next level.
Finally, when reaching the leaf nodes, the binary labels $v_1^N$ corresponding to the edges in path $\{e_1, e_2, \cdots, e_N\}$ which has the largest metric in the list, are assigned to the estimated source vector $\hat{u}_1^N$.

Let $\mathbb{L}^{\left( i \right)}$ denotes the set of candidate paths corresponding to the level-$i$ of code tree in a SCL decoder.
The $\mathbb{L}^{\left( i \right)}$s are stored and updated in a list structure.
The SCL decoding algorithm with searching width $L$, denoted by SCL($L$), can be described as follows:

(A.1) Initialization. A null path is included in the initial list and its metric is set to zero, i.e. ${{\mathbb{L}}^{\left( 0 \right)}}=\left\{ \phi  \right\}$, $M\left( \phi  \right)=0$.

(A.2) Expansion. At the $i$-th level of the code tree, the number of candidate paths in the list are doubled by concatenating new bits ${{v}_{i}}$ taking values of $0$ and $1$ respectively, that is,
\begin{equation}
\mathbb{L}_{{}}^{\left( i \right)}=\left\{ \left( v_{1}^{i-1},{{v}_{i}} \right)\left| v_{1}^{i-1}\in \mathbb{L}_{{}}^{\left( i-1 \right)},{{v}_{i}}\in \left\{ 0,1 \right\} \right. \right\}
\end{equation}

\noindent for each $v_{1}^{i}\in {{\mathbb{L}}^{\left( i \right)}}$, the corresponding path metric(s) are updated according to (\ref{equ_metric_info_or_froz}), (\ref{equ_metric_odd}) and (\ref{equ_metric_even}).

(A.3) Competition. If the number of paths in the list after (A.2) is no more than $L$, just skip this step; otherwise, reserve the $L$ paths with the largest metrics and delete the others.

(A.4) Determination. Repeat (A.2) and (A.3) until level-$N$ is reached.
Then, the decoder outputs the estimated source vector $\hat{u}_1^N=v_1^N$, where $v_1^N$ is the binary labels of the path with the largest metric in the list.

Fig. \ref{fig_scl} gives a simple example of the tree searching under SCL decoding with $L=2$.
Compare with SC in Fig. \ref{fig_tree}, SCL find the most probable path $1000$.
But the times of metric computations is increased from four to seven.

%

SCL maintains $L$ decoding paths simultaneously, each path consumes a $O(N)$ space, the space complexity of SCL then is $O(LN)$.
During the decoding process at each level, each of the $L$ candidates is copied once and extended to two new paths, these copy operations require $O(LN)$ computations.
Moreover, since the code tree has $N$ levels, a direct implementation of SCL decoder will take $O(L N^2)$ computations.
In \cite{SCL:Tal}, a so called ``lazy copy'' technique based on the memory sharing structure among the candidate paths is introduced to reduce this copy complexity.
Therefore, the SCL decoder can be implemented with computational complexity $O(LN\log N)$.

\subsection{Successive Cancellation Stack Decoding}

\begin{figure}[!t]
  \centering
  \includegraphics[width=0.8\columnwidth]{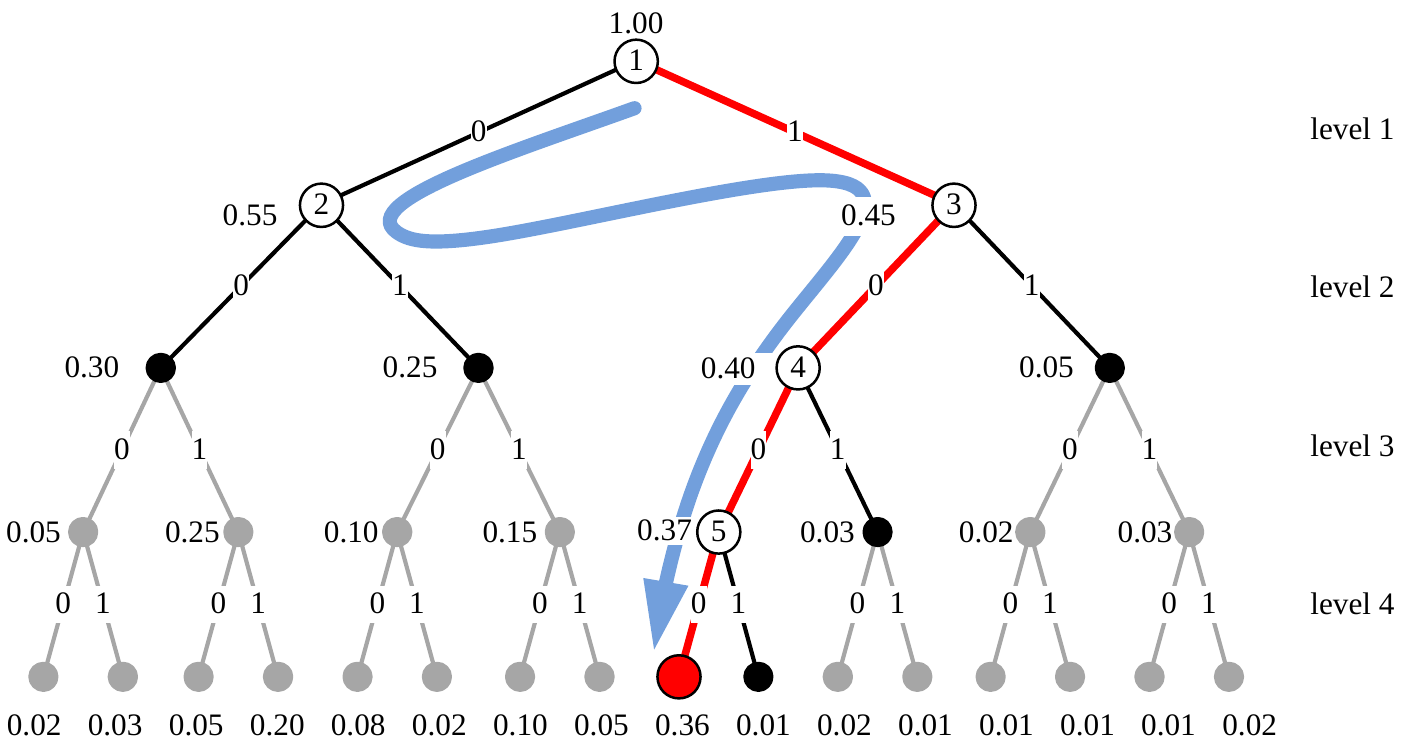}
  \caption{An example of SCS decoding. }
  \label{fig_scs}
\end{figure}

Note that, the path metric (\ref{equ_metric_info_or_froz}) of a certain decoding path with binary label vector $v_1^i$ will not be smaller than that of any of its descendants, i.e.
for any ${v}_{1}^{j} \in \left\{{v}_1^j |  {v}_{i+1}^{j} \in \{0,1\}^{j-i+1} \right\}$ and $i \le j \le N$,
\begin{equation}
M_{N}^{\left( i \right)}\left( v_{1}^{i}\left| y_{1}^{N} \right. \right) \ge {M_{N}^{\left( j \right)}\left( {v}_{1}^{j}\left| y_{1}^{N} \right. \right)}
\end{equation}

Hence, if the metric of a $N$-length decoding path is larger than that of another path with length $l<N$, it must also be larger than the metric of any of the $N$-length descendant path of the latter.
So rather than waiting after processing at each level, we can keep on searching along the single candidate path until its metric is no longer the largest.
Once a $N$-length path is found with the largest metric among all the candidate paths, its binary label vector is simply output as the final estimation, the unnecessary computations for extending other paths are then saved.

The SCS decoder \cite{SCS} uses a ordered stack $\mathbb{S}$ to store the candidate paths and tries to find the optimal estimation by searching along the best candidate in the stack. Whenever the top path in the stack which has the largest path metric reaches length $N$, the decoding process stops and outputs this path. Unlike the candidate paths in the list of SCL which always have the same length, the candidates in the stack of SCS have difference lengths.

Let $D$ denote the maximal the stack $\mathbb{S}$ in SCS decoder.
A little different from the original SCS in \cite{SCS}, an additional parameter $L$ is introduced to limit the number of extending paths with certain length in decoding process.
A counting vector $c_{1}^{N}=\left( {{c}_{1}},{{c}_{2}},...,{{c}_{N}} \right)$ is used to record the number of the popping paths with specific length, i.e. ${{c}_{i}}$ means the number of popped paths with length-$i$ during the decoding process.

The SCS decoding algorithm with the searching width $L$ and the maximal stack depth $D$, denoted by SCS $\left( L, D \right)$, is summarized as follows:

(B.1) Initialization: Push the null path into stack and set the corresponding metric $M\left( \phi  \right)=0$. Initialize the counting vector $s_{1}^{N}$ with all-zeros, and the instantaneous stack depth $|\mathbb{S}|=1$.

(B.2) Popping: Pop the path $v_{1}^{i-1}$ from the top of stack, and if the path is not null, set ${{c}_{i-1}}={{c}_{i-1}}+1$.

(B.3) Expansion: If ${{v}_{i}}$ is a frozen bit, i.e. $i\in{\mathbb{F}}$, simply extend the path to $v_{1}^{i}=\left( v_{1}^{i-1},u_i \right)$; otherwise, if ${{v}_{i}}$ is an information bit, extend current path to $\left( v_{1}^{i-1},0 \right)$ and $\left( v_{1}^{i-1},1 \right)$. Then calculate path metric(s) by (\ref{equ_metric_info_or_froz}), (\ref{equ_metric_odd}) and (\ref{equ_metric_even}).

(B.4) Pushing: For information bit ${{d}_{i}}$, if $|\mathbb{S}|>D-2$, delete the path from the bottom of the stack. Then push the two extended paths into the stack. Otherwise, for frozen bit ${{v}_{i}}$, push the path $v_{1}^{i}=\left( v_{1}^{i-1},0 \right)$ into stack directly.

(B.5) Competition: If ${{c}_{i-1}}=L$, delete all the paths with length less than or equal to $i-1$ from the stack $\mathbb{S}$.

(B.6) Sorting: Resort paths in the stack from top to bottom in descending metrics.

(B.7) Determination: If the top path in the stack reaches to the leaf node of the code tree, pop it from the stack. The decoding algorithm stops and outputs $\hat{u}_1^N=v_1^N$ as the decision sequence. Otherwise go back and execute step (B.2).

Fig. \ref{fig_scs} gives a simple example of the tree searching under SCS.
Compare with SCL in Fig. \ref{fig_scl}, SCS can also find the most probable path $1000$ with two fewer metric computations.

Similar to SC and SCL, the space efficient structure and ``lazy copy'' technique are applied in the implementation of SCS decoders.
The time and space complexity of SCS are $O(LN\log N)$ and $O(DN)$ respectively.
However, under the same searching width $L$, the actual computations of SCS$(L)$ will be much fewer than that of SCL$(L)$ when workding in the moderate or high SNR regime.

\subsection{Hybrid SCL and SCS}

\begin{figure}[!t]
  \centering
  \includegraphics[width=0.8\columnwidth]{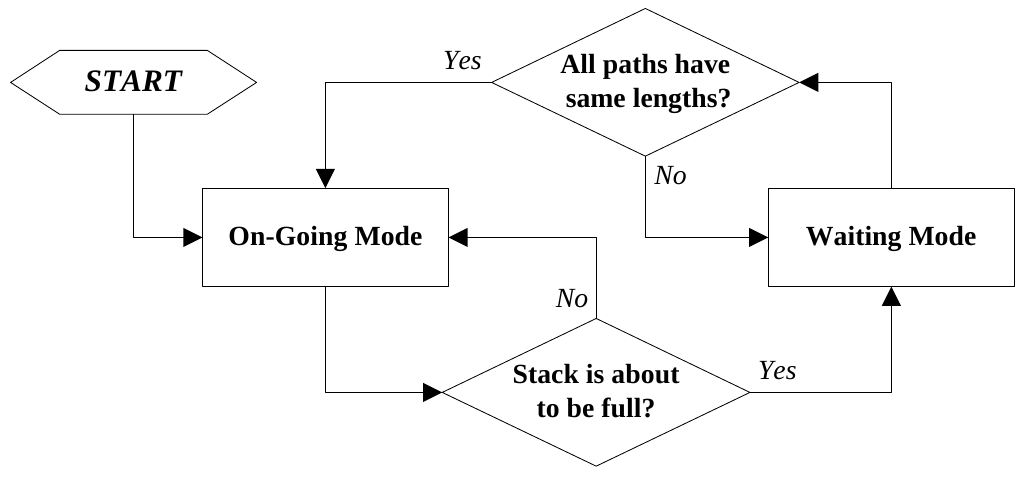}
  \caption{Mode transition diagram of SCH decoding.}
  \label{fig_sch}
\end{figure}

Compared with SCL, SCS decoding can save a lot of unnecessary computations especially when working in the high signal-to-noise (SNR) regime \cite{SCS}.
However, the stack used in SCS consumes a much larger space than SCL.
Theoretically, to prevent performance deterioration, the stack depth $D$ needs to be as large as $LN$, thus the space complexity will becomes $O(LN^2)$.
Fortunately, as shown in \cite{SCS}, a much smaller stack-depth $D$ is enough for moderate and high SNR regimes.
But the most appropriate value of $D$ is relied on the specific SNR and is hard to determine.

In this paper, a new hybrid decoding algorithm called successive cancellation hybrid (SCH) is proposed.
SCH, as the name suggests, is a hybrid of SCL and SCS.
SCH has two working modes called \emph{on-going} and \emph{waiting}.
At first, SCH decoder works on the on-going mode, it searches along the best candidate path using a ordered stack just the same as that SCS does.
But when the stack is about to be full, SCH stops searching forward and switches to the waiting mode.
Under the waiting mode, SCH turns to extend the shortest path in the stack until all the candidate paths in the stack have the same length.
The processing under waiting mode is somewhat similar to SCL and it decreases the number of paths in stack.
Then, SCH switches back to the on-going mode again.
Fig. \ref{fig_sch} gives a graphic illustration.
This decoding procedure goes on until an $N$-length path appears at the top of the stack.

The SCH algorithm with the searching width $L$, the maximal stack depth $D$, denoted by SCH $\left( L,D \right)$, is summarized as follows:

(C.1) Initialization: Push the null path into stack $\mathbb{S}$ and set the corresponding metric $M\left( \phi  \right)=0$. Initialize the counting vector $c_{1}^{N}$ with all-zeros, and the instantaneous stack depth $|\mathbb{S}|=1$. The working mode flag $f_{mode}$ is set to $0$, where $0$ denote the on-going mode and $1$ denote the waiting mode.

(C.2) Popping: When $f_{mode}=0$, pop the path $v_{1}^{i-1}$ from the top of stack; else when $f_{mode}=1$, pop the path $v_{1}^{i-1}$ with the shortest path length in the stack. Then, if the popped path is not null, i.e. $v_{1}^{i-1} \neq \phi$, set ${{c}_{i-1}}={{c}_{i-1}}+1$.

(C.3) Expansion: If ${{v}_{i}}$ is a frozen bit, i.e. $i \in \mathbb{F}$, simply extend the path to $v_{1}^{i}=\left( v_{1}^{i-1}, u_i \right)$; otherwise, if ${{v}_{i}}$ is an information bit, i.e. $i \in \mathbb{I}$, extend current path to $\left( v_{1}^{i-1},0 \right)$ and $\left( v_{1}^{i-1},1 \right)$. Then calculate path metric(s) by (\ref{equ_metric_info_or_froz}), (\ref{equ_metric_odd}) and (\ref{equ_metric_even}).

(C.4) Pushing: For information (frozen) bit ${{v}_{i}}$, push the new two paths (one path) into the stack.

(C.5) Competition: If ${{c}_{i-1}}=L$, delete all the paths with length less than or equal to $i-1$ from the stack $\mathbb{S}$.

(C.6) Mode Switching: When $f_{mode}=0$ and $D-|\mathbb{S}| \le 2L-1$, switch $f_{mode}=1$; when $f_{mode}=1$ and all the candidate pathes in the stack have equal lengths, $f_{mode}=1$;

(C.7) Sorting: Resort paths in the stack from top to bottom in descending metrics.

(C.8) Determination: If the top path in the stack reaches to the leaf node of the code tree, pop it from the stack. The decoding algorithm stops and outputs $\hat{u}_1^N=v_1^N$ as the decision sequence. Otherwise go back and execute step (C.2).

The time and space complexity of SCH are $O(LN\log N)$ and $O(DN)$ respectively.
The actual computations of SCH decoding is less than that of SCL but is usually more than that of SCS.

For SCH decoding, since no path is dropped when the stack is about to be full, the performance will not affected by $D$.
However, when decoder stays in the waiting mode, unnecessary computations will be taken.
And the smaller the maximum stack depth is $D$, the more likely the decoder will switch to the waiting mode.
So, the computational complexity grows with the decreasing of $D$.
To have enough space for waiting mode, the minimum value of $D$ is $2L$.
Particularly, when $D=2L$, SCH($L$,$D$) is equivalent to SCL($L$); and when $D \ge LN$, SCH($L$,$D$) is equivalent to SCS($L$,$D$).

\subsection{Pruning Technique}
\label{subsec_prune}
\begin{figure}[!t]
  \centering
  \includegraphics[width=0.8\columnwidth]{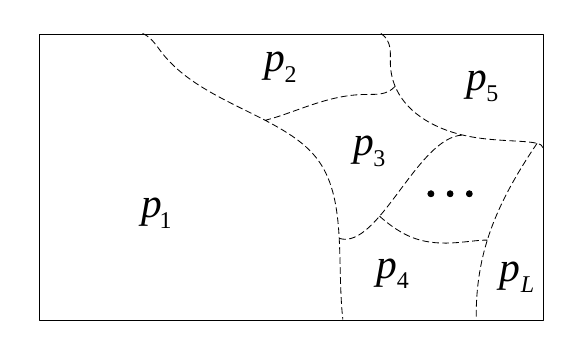}
  \caption{The $L$ candidates divide the probability space into $L$ partitions.}
  \label{fig_prob_space}
\end{figure}

During the path searching on the code tree, the candidate paths with too small metrics and their descendants will hardly have the chance to be reserved in the future process.
In this subsection, we propose a pruning technique to reduce the computational complexity of the improved successive cancellation decoding algorithms.

An additional vector $a_1^N$ is used to record the pruning reference for each level, where $a_i$ is the largest metric of all the traversed $i$-length decoding paths on the code tree.
More specifically, for SCL decoding,
\begin{equation}
\label{equ_max_metric_SCL}
a_i = \max \limits _{v_1^i \in \mathbb{L}^{(i)}} { M_{N}^{\left( i \right)}\left( v_{1}^{i}\left| y_{1}^{N} \right. \right) }
\end{equation}
And equivalently, for SCS and SCH, $a_i$ is set to the metric of the first $i$-length path popped off the stack.

We introduce a new parameter called probability ratio threshold $\tau$.
During the processing at level-$i$ on the code tree, a $i$-length path with metric smaller than $a_i-\log(\tau)$ is dropped directly.
Recall that the path metrics are defined as the logarithmic APPs (\ref{equ_metric_info_or_froz}).
Therefore, the pruned paths are those whose APPs
\begin{equation}
P_{N}^{\left( i \right)}\left( v_{1}^{i}\left| y_{1}^{N} \right. \right) < \frac {\exp \left(a_i\right)}{\tau}
\end{equation}

Intuitively, the correct path will possibly be dropped in this pruning operation.
In the following part of this subsection, an upperbound of the additional performance deterioration brought by $\tau$ is derived and a conservative configuration of $\tau$ is given.

Hereafter, SCL, SCS and SCH are collectively referred to as improved successive cancellation (ISC) decoding algorithms.
The block error event of polar code with information set $\mathbb{I}$ under ISC decoding is defined as
\begin{equation}
\mathcal{E} = \left\{ \left(u_1^N, \hat{u}_1^N, y_1^N\right) \in \mathbb{X}^N \times \mathbb{X}^N \times \mathbb{Y}^N:  u_{\mathbb{I}} \neq \hat{u}_{\mathbb{I}}\right\}
\end{equation}

By introducing pruning operations, the error events can be classified into two kinds.
The first kind is the correct path is not lost until the final decision phase, i.e. the correct path is contained in the final list(or stack) but does not have the largest metric.
The second kind is the correct path is lost before the decision step.
So, the block error rate (BLER) of ISC can be decomposed as
\begin{equation}
P_{ISC}{\left(N, \mathbb{I}, L, D, \tau\right)} = P(\mathcal{E}|\overline{\mathcal{C}})P(\overline{\mathcal{C}})+P(\mathcal{C})
\end{equation}
where ${\mathcal{C}}$ means the correct path loss.

The event $\mathcal{C}$ can be further decomposed as
\begin{equation}
P(\mathcal{C}) = \sum \nolimits_{i \in \mathbb{I}} P(\mathcal{C}_i)
\end{equation}
where $\mathcal{C}_i$ is the event that the correct path is not lost until the processing at the $i$-th level.

There are three kinds of event which will lead to path loss at the $i$-th level.
The first is brought by the searching width limitation, i.e. the correct path is excluded from the $L$ best paths in $i$-th decoding step, and is denoted by $\mathcal{L}_i$.
The second is brought by the maximum probability ratio limitation, i.e. the metric of the correct path is much smaller than that of the best one, and is denoted by $\mathcal{T}_i$.
The third is brought by the maximum stack depth limitation, which only exist in the SCS decoding that the correct path is abandoned when the path length equals $i$ and the metric is much smaller than the maximum one at that moment, and this event is denoted by $\mathcal{D}_i$.
Then
\begin{eqnarray}
\label{equ_event_ci}
P(\mathcal{C}_i)=P(\mathcal{L}_i)+P(\mathcal{T}_i|\overline{\mathcal{L}}_i)P(\overline{\mathcal{L}}_i)
+P(\mathcal{D}_i|\overline{\mathcal{L}}_i \overline{\mathcal{T}}_i)P(\overline{{\mathcal{L}}}_i \overline{{\mathcal{T}}}_i)
\end{eqnarray}
For SCL, SCH decoding or SCS with a large enough stack depth $D$, $P(\mathcal{D}_i|\overline{\mathcal{L}}_i \overline{\mathcal{T}}_i)=0$.

The additional BLER performance deterioration brought by pruning is
\begin{equation}
\sum \nolimits_{i \in \mathbb{I}} P(\overline{\mathcal{L}}_i \mathcal{T}_i ) =
\sum \nolimits_{i \in \mathbb{I}} P(\mathcal{T}_i|\overline{\mathcal{L}}_i)P(\overline{\mathcal{L}}_i)
\end{equation}

During the processing on the code tree $\mathbb{T}$ , we will have at most $L$ paths at level-$i$ with APPs $\{p_1, p_2, \cdots, p_L\}$ which is calculated by (\ref{equ_app}),
and
\begin{equation}
q = \sum \limits _{j=1}^{L}{p_j}\le 1
\end{equation}
Without loss of generality, we assume that $p_1 \ge p_2 \ge \cdots \ge p_L$.
By the assumption that the one of these paths is the correct path, the $L$ probability divided the whole probability space into $L$ parts as shown in Fig. \ref{fig_prob_space}.
The event of correct path loss in the pruning processing at the $i$-th level has a probability
\begin{equation}
P(\mathcal{T}_i | \overline{\mathcal{L}}_i)=\frac{1}{q} \sum \limits_{j\in\{2,3,\cdots, L\}, p_j<p_1} p_j
\end{equation}

For each of these eliminated paths, the corresponding probability
\begin{equation}
\label{equ_ub_pi}
p_j \le \frac{p_1}{\tau} \le \frac{q}{\tau}
\end{equation}
where $j\in\{2, 3,\cdots, L\}$.
So we have
\begin{equation}
P(\mathcal{T}_i| \overline{\mathcal{L}}_i) \le \frac{L-1}{\tau}
\end{equation}

The additional error probability brought by $\tau$ is upper bounder by
\begin{equation}
\label{equ_ptau_upperbound}
\sum \nolimits_{i \in \mathbb{I}} P(\overline{\mathcal{L}}_i \mathcal{T}_i ) \le \sum \nolimits_{i \in \mathbb{I}} P(\mathcal{T}_i| \overline{\mathcal{L}}_i) \le \frac{K(L-1)}{\tau}
\end{equation}

Given a tolerable performance deterioration $P_{tol}$, the value of $\tau$ can be determined as
\begin{equation}
\label{equ_tau_value}
{\tau} = \frac{K(L-1)}{P_{tol}}
\end{equation}

In most cases, since the upperbound in (\ref{equ_ptau_upperbound}) is very loose, the accrual performance deterioration is usually far less than $P_{tol}$.
The configuration of $\tau$ in (\ref{equ_tau_value}) is very conservative.

\section{Simulation Results}
\label{section_simulations}

In this section, the performance and complexity of the improved successive cancellation (ISC) decoding algorithms will be discussed.

To simplify the complexity evaluation of polar decoding, we measure the average computational complexity in terms of the number of metric recursive operations, which are defined in (\ref{equ_metric_odd}) or (\ref{equ_metric_even}).
For example, the computational complexity of SC decoder is $N\log N=1024\times 10\approx {{10}^{4}}$.

\begin{figure}[!t]
  \centering
  \includegraphics[width=0.8\columnwidth]{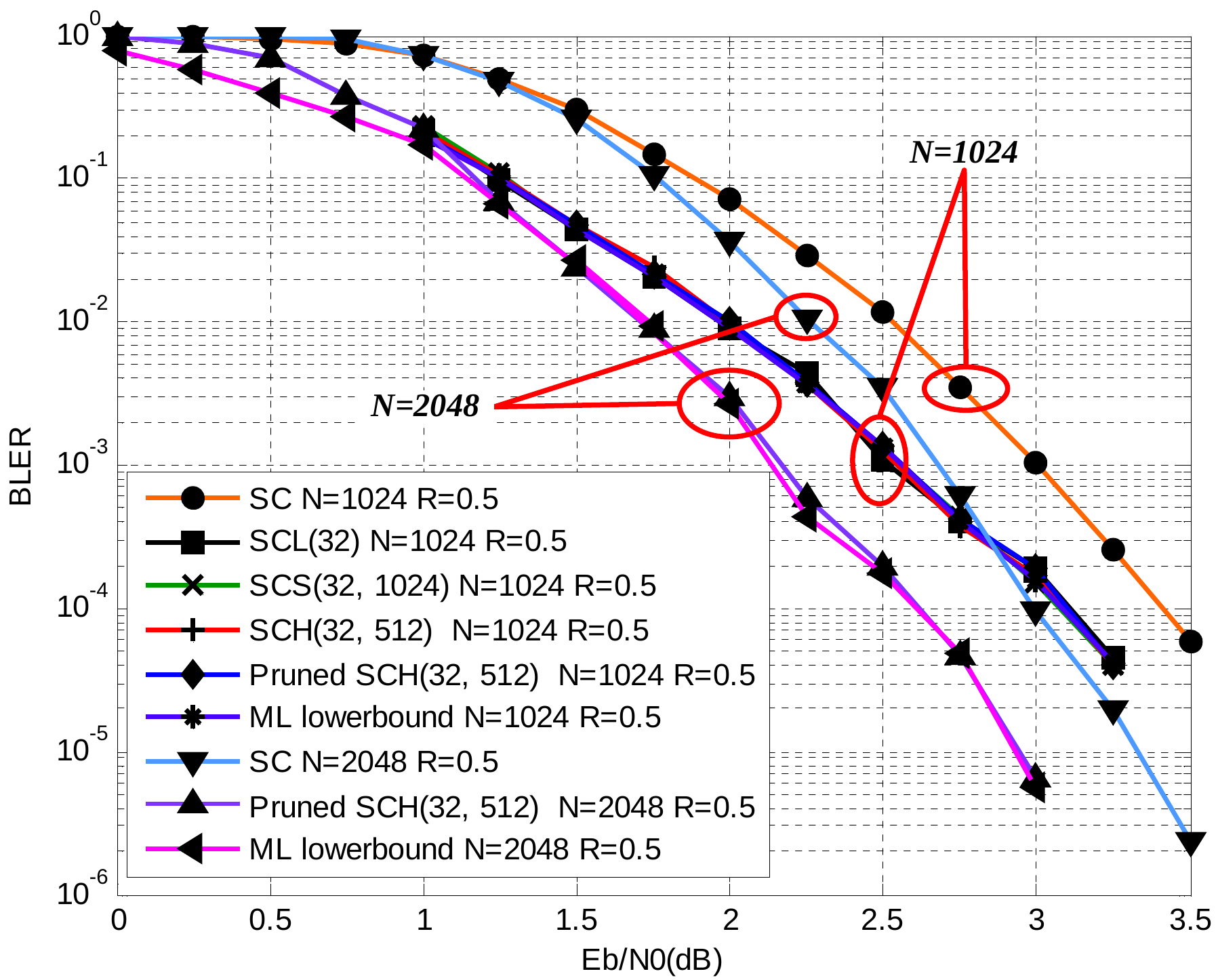}
  \caption{BLER under different code length}
  \label{fig_diff_len_performance}
\end{figure}

\begin{figure}[!t]
  \centering
  \includegraphics[width=0.8\columnwidth]{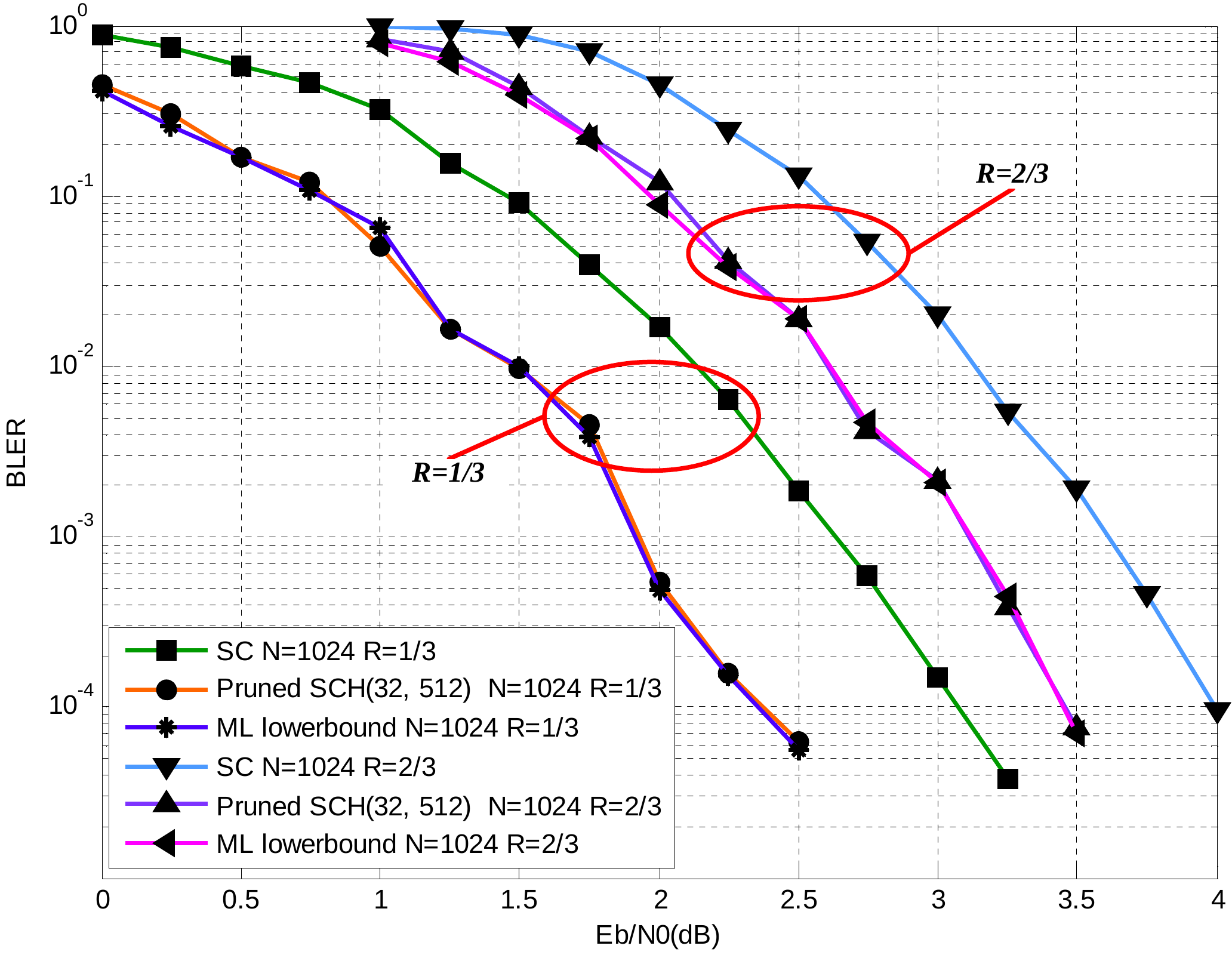}
  \caption{BLER under different code rate}
  \label{fig_diff_rate_performance}
\end{figure}

\begin{figure}[!t]
  \centering
  \includegraphics[width=0.8\columnwidth]{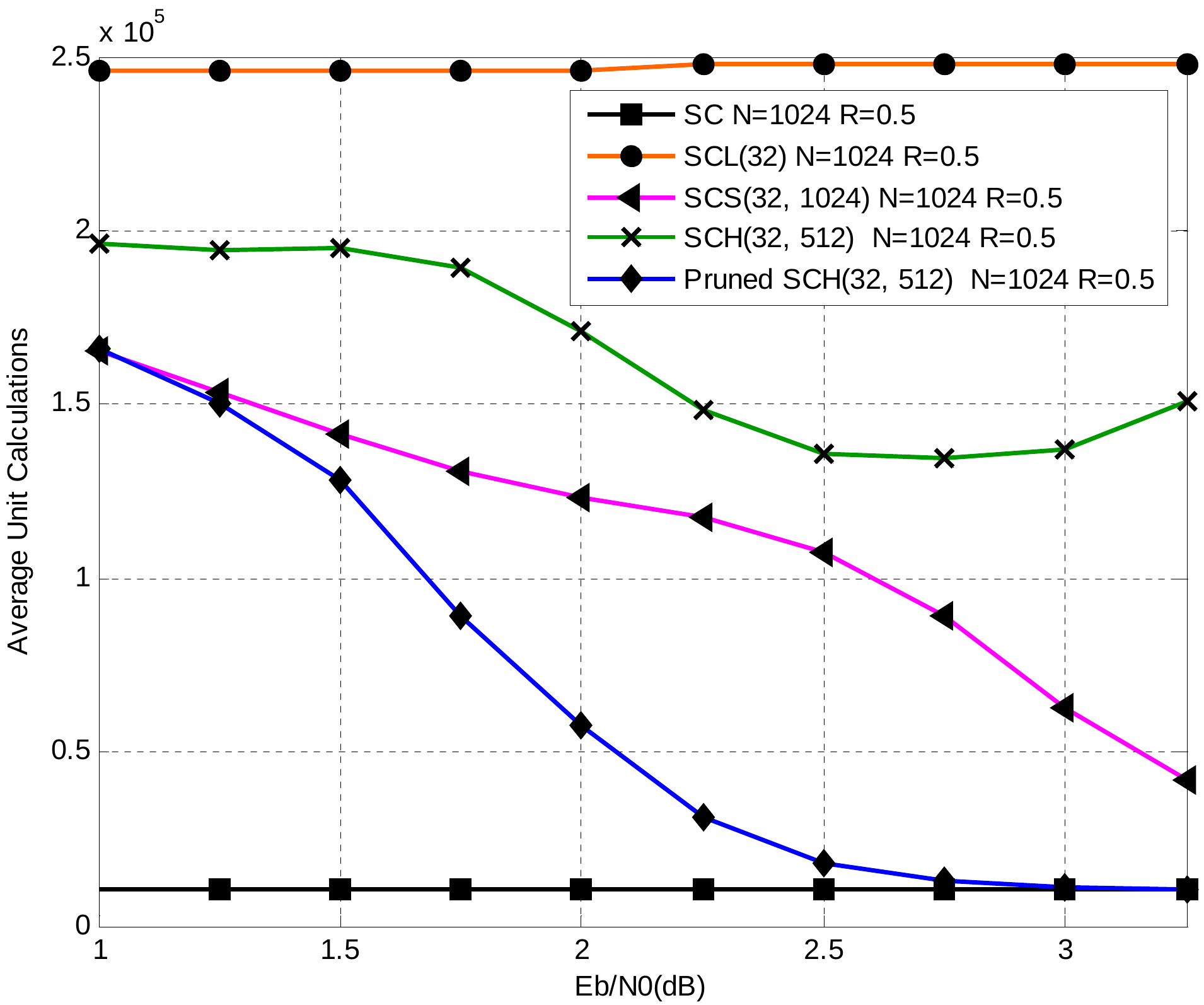}
  \caption{Complexity under different decoding algorithms}
  \label{fig_diff_dec_complexity}
\end{figure}

Fig. \ref{fig_diff_len_performance} gives the simulation results with code length $N$ set as $1024$ and $2048$,  and the code rate $R=1/2$.
And Fig. \ref{fig_diff_rate_performance} shows the BLER performances with code rate $R$ set as $1/3$ and $2/3$, and the code length $N$ is fixed to $1024$.
The lowerbounds of BLER performance under maximum-likelihood (ML) decoding are obtained by performing SCL($32$) decoding and counting the number of times the decoded codeword is more likely than the transmitted one.
The probability ratio threshold $\tau$ for pruning operation is set by (\ref{equ_tau_value}) with $P_{tol}=10^{-5}$.
As shown in the figures, under proper configurations, all the three decoding algorithms can achieve the performance very close to that of ML decoding.

The average computational complexities under different decoding algorithms with code length $N=1024$ and code rate $R=1/2$ are shown in Fig. \ref{fig_diff_dec_complexity}.
We can see that the complexity of SCH is not monotonically decreasing with the increasing of SNR.
This is because the switching between the two working modes is relied on the certain code construction and searching procedures.
However, SCH always has a much lower computational complexity than that of SCL.
Although it needs more computations than SCS, SCH occupies less memory space without any deterioration in performance.
In fact, under some specific configurations, SCH can be equivalent to the other two decoding algorithms:
when $D=2L$, SCH($L$, $2L$) is equivalent to SCL($L$);
and when $D$ is very large, SCH($L$, $D$) is equivalent to SCS($L$, $D$);
Therefore, SCH can achieve a better trade-off between computational complexity and space complexity.
Furthermore, by applying the pruning technique introduced in section \ref{subsec_prune}, the computational complexity can be significantly reduced and very close to that of SC in the moderate and high SNR regime.

Compared with SC, ISC decoding algorithms introduce three more parameters: the searching width $L$, the maximum stack depth $D$ and probability ratio threshold for pruning $\tau$.
In the following part of this section, we will analysis the impacts on performance and complexity of this three parameters one-by-one.

\subsection{On Different Searching Width $L$}

\begin{figure}[!t]
  \centering
  \includegraphics[width=0.8\columnwidth]{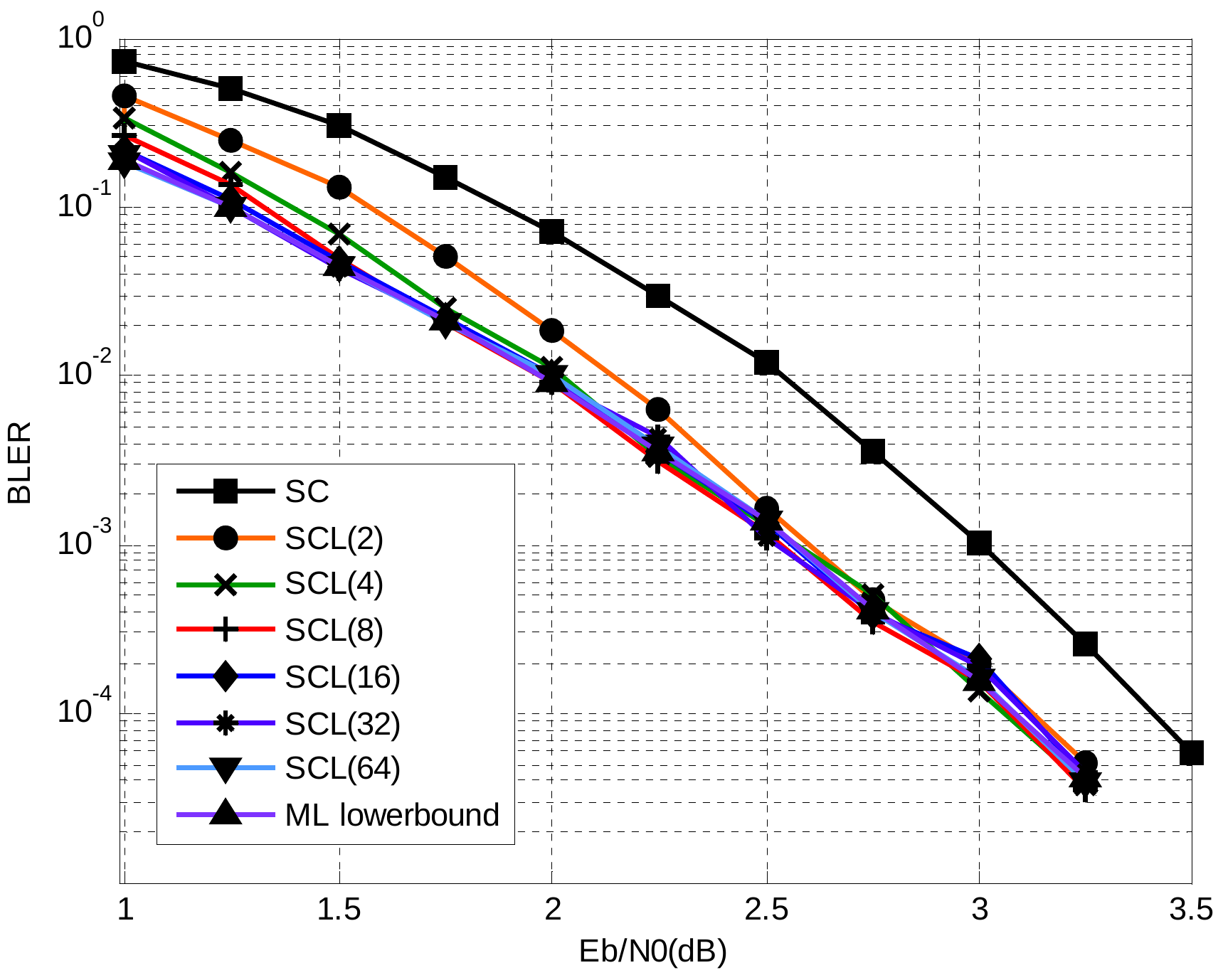}
  \caption{BLER under different $L$}
  \label{fig_diff_L_performance}
\end{figure}

\begin{figure}[!t]
  \centering
  \includegraphics[width=0.8\columnwidth]{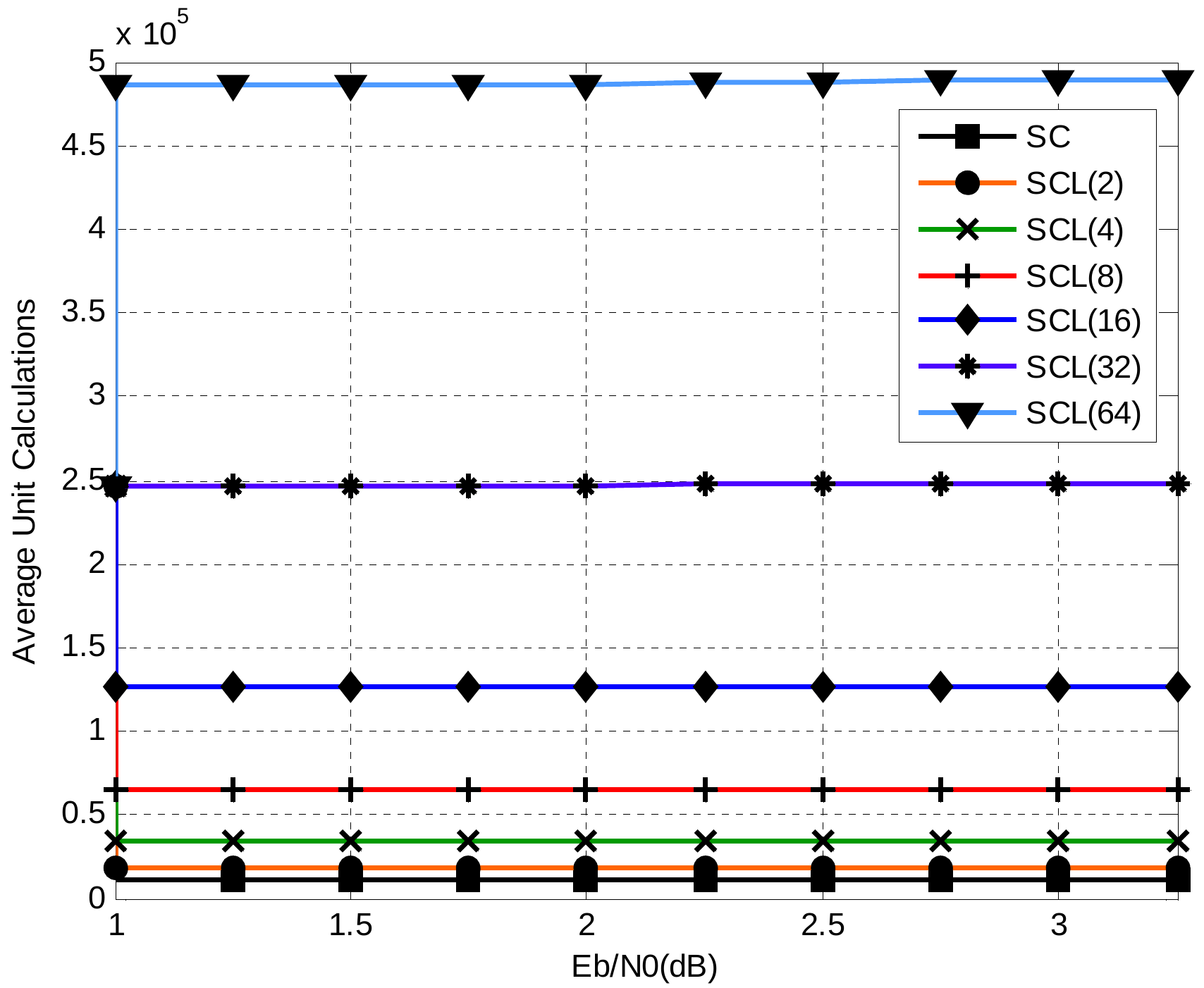}
  \caption{Complexity under different $L$}
  \label{fig_diff_L_complexity}
\end{figure}

Fig. \ref{fig_diff_L_performance} gives the performance comparisons under SCL decoding with different $L$.
The code length and code rate are set as $N=1024$ and $R=0.5$ respectively.
The searching width $L$ varies from $1$ (equivalent to SC) to $64$.

Note that, SCL($L$) is equivalent to SCH($L$, $2L$) and SCS($L$, $D$) with a large enough $D$.
The affects brought by different $L$ in SCL decoding are the same with that in SCS and SCH.

The larger the searching width is, the less probable to lose the correct path, i.e. $P(\mathcal{L}_i)$ in (\ref{equ_event_ci}) is a decreasing function of $L$.
But according to the results depicted in Fig. \ref{fig_diff_L_complexity}, the computational complexity is approximately proportional to $L$.
As shown in Fig. \ref{fig_diff_L_performance}, $L=32$ is good enough for $N=1024$ and $R=0.5$.

\subsection{On Different Stack Depth $D$}

\begin{figure}[!t]
  \centering
  \includegraphics[width=0.8\columnwidth]{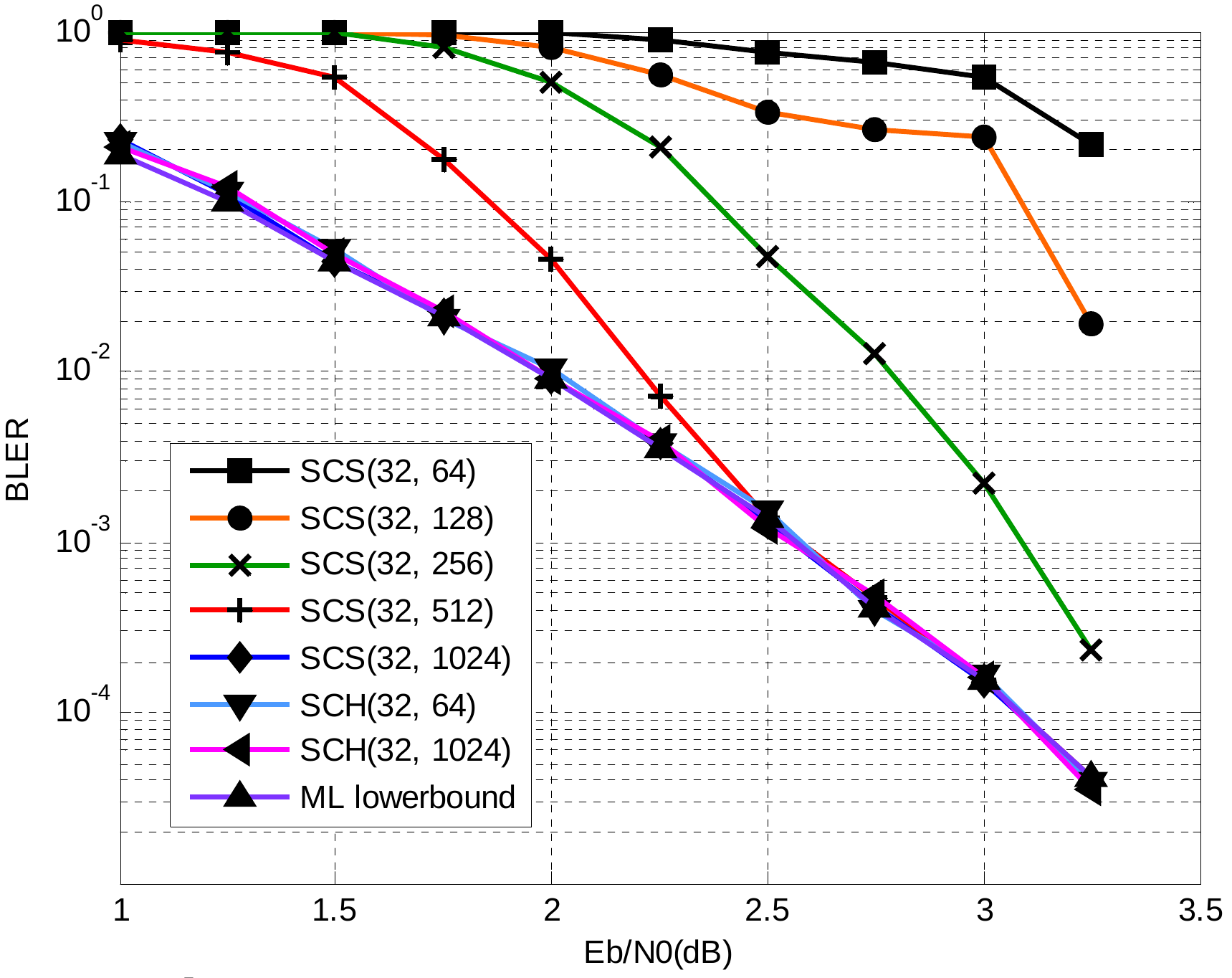}
  \caption{BLER under different $D$}
  \label{fig_diff_D_performance}
\end{figure}

\begin{figure}[!t]
  \centering
  \includegraphics[width=0.8\columnwidth]{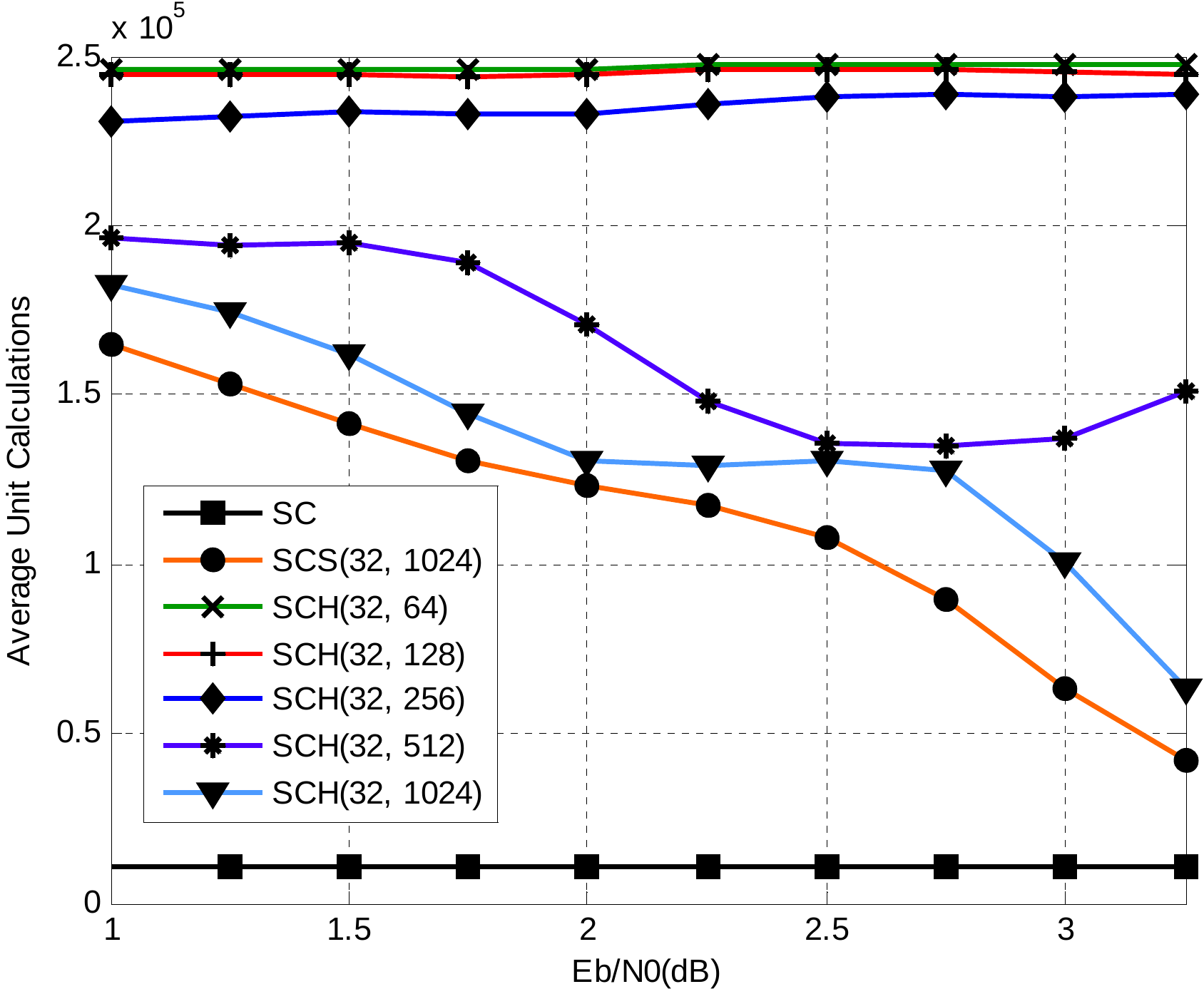}
  \caption{Complexity under different $D$}
  \label{fig_diff_D_complexity}
\end{figure}

For polar codes under SCS decoding, a too small value of the maximum stack depth $D$ will lead to significant deterioration on performance.
As shown in Fig. \ref{fig_diff_D_performance}, $D$ need to be larger than $1024$ for SCS decoding.
But for SCH, the different configurations of $D$ no longer affect the BLER performance but the computational complexity.
As shown in Fig. \ref{fig_diff_D_complexity}, the computational complexity of SCH is decreasing with the increasing of $D$.
Although it needs more computations than SCS, SCH occupies less memory space without any deterioration in performance.
Compared with SCL, SCH has much lower computational complexity and only require a little more memory space.
In fact, under some specific configurations, SCH can be equivalent to the other two decoding algorithms:
when $D=2L$, SCH($L$, $2L$) is equivalent to SCL($L$);
and when $D$ is very large, SCH($L$, $D$) is equivalent to SCS($L$, $D$);
Hence, SCH can achieve a better trade-off between computational complexity and space complexity.

\subsection{On Different Pruning Ratio $\tau$}

\begin{figure}[!t]
  \centering
  \includegraphics[width=0.8\columnwidth]{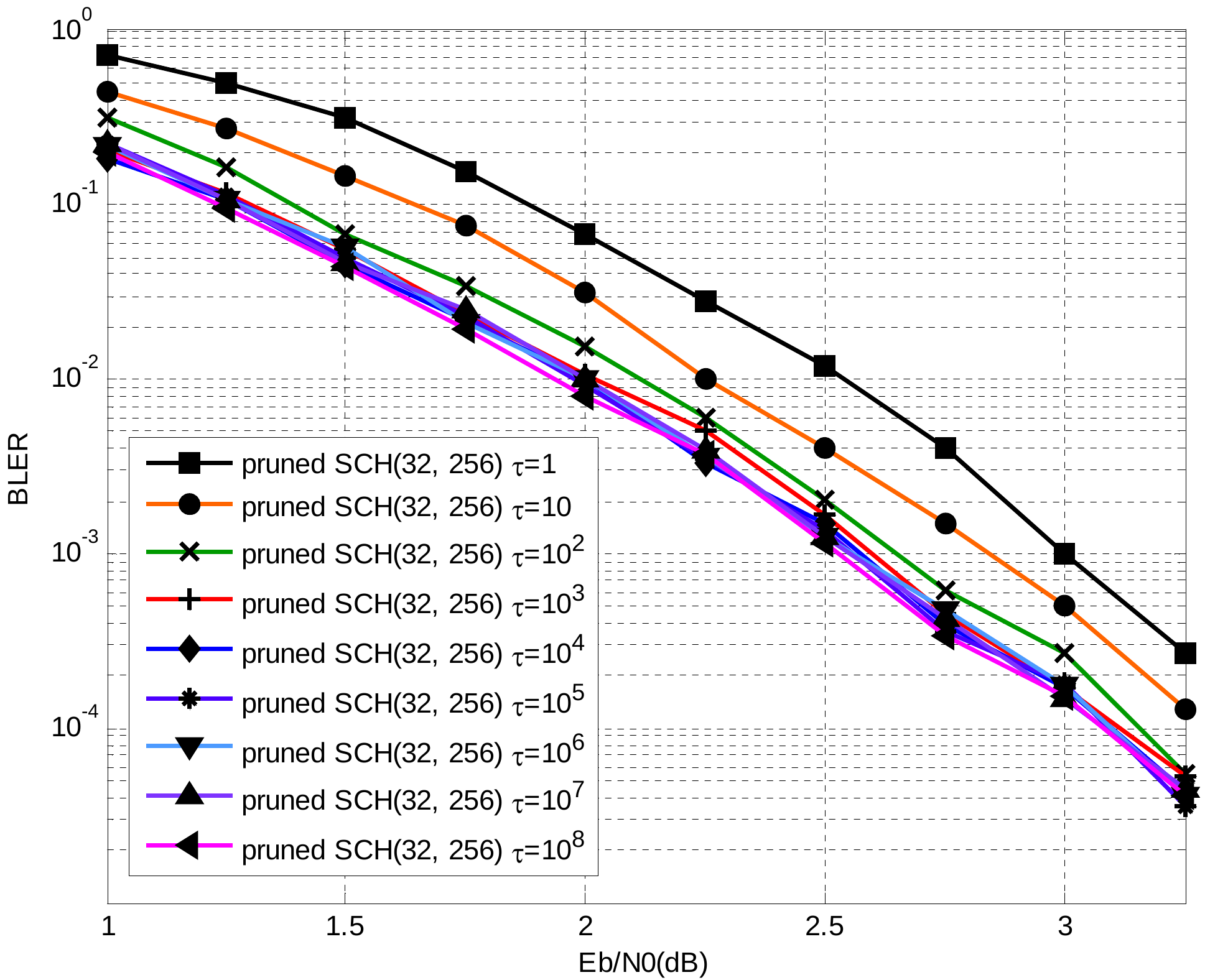}
  \caption{BLER under different $\tau$}
  \label{fig_diff_tau_performance}
\end{figure}

\begin{figure}[!t]
  \centering
  \includegraphics[width=0.8\columnwidth]{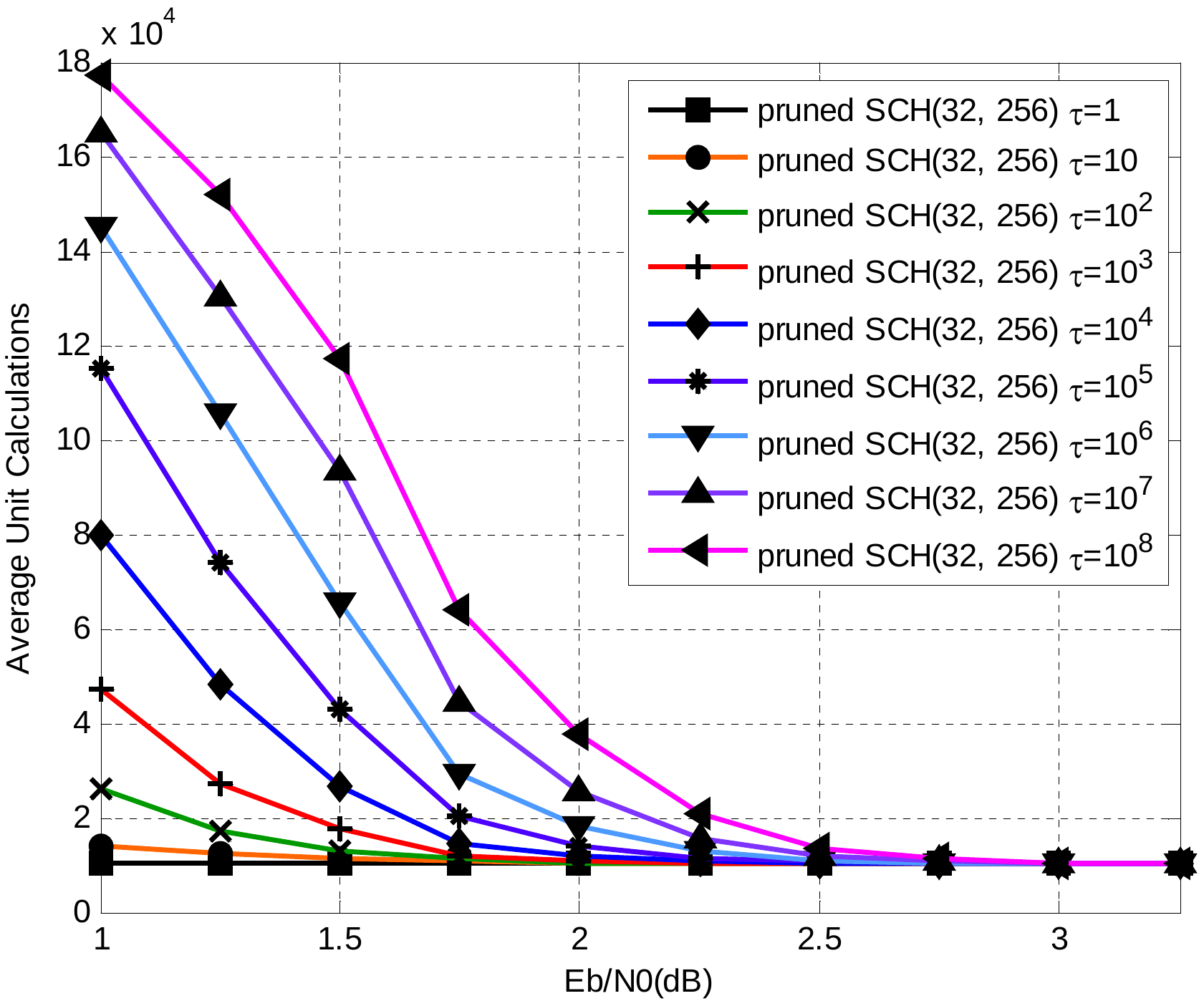}
  \caption{Complexity under different $\tau$}
  \label{fig_diff_tau_complexity}
\end{figure}

Fig. \ref{fig_diff_tau_performance} and Fig. \ref{fig_diff_tau_complexity} give simulations of polar codes with code length $N=1024$ code rate $R=0.5$ over binary-input additive Gaussian noise channels (BAWGNCs).
The codes are decoded by SCH decoders with $L=32$, $D=256$ and $\tau$ varies from $1$ to $10^8$.
As shown in the figures, the computational complexity will be reduced when the increasing of $\tau$, while the BLER performance will be deteriorated with a too small $\tau$.
Larger values of $\tau$ such as $10^4\sim10^8$ will introduce little deterioration in performance, but will lead to larger complexities.
However, when the codes work in a moderate signal-to-noise ratio (SNR) regime such as $2.5$dB where the BLER is less than $10^{-3}$, the computational complexity differences of SC and SCH decoding under different $\tau$ in the simulated regime tends to negligible as shown in Fig. \ref{fig_diff_tau_complexity}.




\section{Conclusion}
\label{section_conclusion}
The successive cancellation (SC) decoding algorithm of polar codes and its improved versions, successive cancellation list (SCL) and successive cancellation stack (SCS) are restated as path searching procedures on the code tree of polar codes.
Combining the ideas of SCL and SCS, a new decoding algorithm named successive cancellation hybrid (SCH) is proposed, which can achieve a better trade-off between computational complexity and space complexity.
To avoid unnecessary path searching, a pruning technique which is suitable for all improved successive cancellation (ISC) decoders is proposed.
Performance and complexity analysis based on simulations show that, with the help of the pruning technique, all the ISC decoders can have a performance very close to that of maximum-likelihood (ML) decoding, and the computational complexities can be very close to that of SC in the moderate and high signal-to-noise ratio (SNR) regime.

%

\ifCLASSOPTIONcaptionsoff
  \newpage
\fi


\begin{thebibliography}{1}

\bibitem{Arikan}
E.~Arikan, ``Channel Polarization: A Method for Constructing Capacity-Achieving Codes for Symmetric Binary-Input Memoryless Channels,'' \emph{IEEE Trans. Inf. Theory}, vol. 55, no. 7, pp. 3051-3073, Jul. 2009.

\bibitem{rate}
E. Ar{\i}kan, and E. Telatar, ``On the rate of channel polarization,`` \emph{IEEE Int. Symp. Inform. Theory (ISIT)}, pp. 1493-1495, Jul. 2009.

\bibitem{DE}
R.~Mori and T.~Tanaka, ``Performance of polar codes with the construction using density evolution,'' \emph{IEEE Commun. Lett.}, vol. 13, no. 7, pp. 519-521, Jul. 2009.


\bibitem{characterization}
S. B. Korada, E. Sasoglu, and R. Urbanke, ``Polar Codes: Characterization of Exponent, Bounds, and Constructions,'' \emph{IEEE Trans. Inf. Theory}, vol. 56, no. 12, pp. 6253-6264, 2010.


\bibitem{construct}
I. Tal and A. Vardy, ``How to construct polar codes,'' arXiv:1105.6164v1, May 2011.

\bibitem{construct_on}
R.~Pedarsani, S.~H.~Hassani, I.~Tal, and E.~Telatar, ``On the construction of polar codes,'' \emph{IEEE Int. Symp. Inform. Theory (ISIT)}, pp. 11-15, Jul. 2011.

\bibitem{constructq}
S. Cayci, O. Arikan, and E. Arikan, ``Polar code construction for non-binary source alphabets,'' \emph{20th Signal Processing and Communications Applications Conference (SIU)}, pp. 1-4, 2012.

\bibitem{RMdec}
I. Dumer and K. Shabunov, ''Recursive and pennutation decoding for Reed-Muller codes,'' \emph{IEEE Int. Symp. Inform. Theory (ISIT)}, pp. 146, 2002.

\bibitem{RMlist}
I. Dumer and K. Shabunov, ``Soft-decision decoding of Reed-Muller codes: recursive lists,'' \emph{IEEE Trans. Inf. Theory}, vol. 52, no. 3, pp. 1260-1266, Mar. 2006.

\bibitem{RMstack}
N. Stolte, U. Sorger, and G. Sessler, ``Sequential stack decoding of binary Reed-Muller codes,'' in 3rd ITG Conference Source and Channel Coding, Jan. 2000.

\bibitem{SCL:Tal}
I.~Tal and A.~Vardy, ``List decoding of polar codes,'' \emph{IEEE Int. Symp. Inform. Theory (ISIT)}, pp. 1-5, 2011.

\bibitem{SCL}
K.~Chen, K.~Niu, and J.~R. Lin, ``List successive cancellation decoding of polar codes,'' \emph{Electronics Letters}, vol. 48, no. 9, pp. 500-501, 2012.

\bibitem{SCS}
K.~Niu and K.~Chen, ``Stack decoding of polar codes,'' \emph{Electronics Letters}, vol. 48, no. 12, pp. 695-696, 2012.


\bibitem{parallel}
E. Hof, I. Sason, and S. Shamai, ``Polar coding for reliable communications over parallel channels,'' \emph{IEEE Information Theory Workshop (ITW)}, pp. 1-5, Aug. 2010.


\bibitem{perform}
N. Hussami, S. B. Korada, and R. Urbanke, ``Performance of polar codes for channel and source coding,'' \emph{IEEE Int. Symp. Inform. Theory (ISIT)}, pp. 1488-1492, Jul. 2009.

\bibitem{LP}
N. Goela, S. B. Korada, and M. Gastpar, ``On LP Decoding of Polar Codes,'' \emph{IEEE Information Theory Workshop (ITW)}, pp. 1-5, Aug. 2010.

\bibitem{ML}
E. Ar{\i}kan, H. Kim, G. Markarian, U. Ozgur, and E. Poyraz, ``Performance of short polar codes under ml decoding,'' \emph{ICT-Mobile Summit 2009 Conference Proc.}, 2009.

\bibitem{multilevel}
W. Park, and A. Barg, ``Multilevel polarization for nonbinary codes and parallel channels,'' \emph{49th Annual Allerton Conference on Communication, Control, and Computing (Allerton)}, pp. 228-234, 2011.

\bibitem{pocm}
D. M. Shin, S. C. Lim, and K. Yang, ``Mapping Selection and Code Construction for $2^m$-ary Polar-Coded Modulation,'' \emph{IEEE Commun. Lett.}, vol. 16, no. 6, pp. 905-908, 2012.

\bibitem{mac1}
E. \c{S}a\c{s}o\u{g}lu, E. Telatar, and E. Yeh, ``Polar codes for the two-user binary-input multiple-access channel,'' \emph{IEEE Information Theory Workshop (ITW)}, pp. 1-5, 2010.

\bibitem{mac2}
E. Abbe, and E. Telatar, ``MAC polar codes and matroids,'' \emph{Information Theory and Applications Workshop (ITA)}, pp. 1-8, 2010.

\bibitem{source_coding1}
S. B. Korada, and R. L. Urbanke, ``Polar Codes are Optimal for Lossy Source Coding,'' \emph{IEEE Trans. Inf. Theory}, vol. 56, no. 4, pp. 1751-1768, 2010.

\bibitem{security1}
E. Koyluoglu, and H. El Gamal, ``Polar coding for secure transmission and key agreement,'' \emph{Proc. IEEE Int. Symp. Personal, Indoor and Mobile Radio Commun. Conf (PIMRC)}, pp. 2698-2703, 2010.

\bibitem{security2}
E. Hof, and S. Shamai, ``Secrecy-achieving polar-coding,'' \emph{IEEE Information Theory Workshop (ITW)}, pp. 1-5, 2010.


\end{thebibliography}
\end{document}